\documentstyle[a4,12pt]{article}

\newcommand{\be}{\begin{equation}}
\newcommand{\ee}{\end{equation}}
\newcommand{\ba}{\begin{eqnarray}}
\newcommand{\ea}{\end{eqnarray}}
\newcommand{\tfrac}[2]{{\textstyle{#1\over#2}}}
\newcommand{\ys}{
\setlength{\unitlength}{1.0pt}
\begin{picture}(30,10)(0,0)
\thinlines
\put ( 0, 0) {\line (1,0){30}}
\put ( 0,10) {\line (1,0){30}}
\put ( 0, 0) {\line (0,1){10}}
\put (10, 0) {\line (0,1){10}}
\put (20, 0) {\line (0,1){10}}
\put (30, 0) {\line (0,1){10}}
\put ( 3, 2) {1}
\put (13, 2) {2}
\put (23, 2) {3}
\end{picture}}
\newcommand{\yml}{
\setlength{\unitlength}{1.0pt}
\begin{picture}(20,20)(0,5)
\thinlines
\put ( 0, 0) {\line (1,0){10}}
\put ( 0,10) {\line (1,0){20}}
\put ( 0,20) {\line (1,0){20}}
\put ( 0, 0) {\line (0,1){20}}
\put (10, 0) {\line (0,1){20}}
\put (20,10) {\line (0,1){10}}
\put ( 3, 2) {3}
\put ( 3,12) {1}
\put (13,12) {2}
\end{picture}}
\newcommand{\ymr}{
\setlength{\unitlength}{1.0pt}
\begin{picture}(20,20)(0,5)
\thinlines
\put ( 0, 0) {\line (1,0){10}}
\put ( 0,10) {\line (1,0){20}}
\put ( 0,20) {\line (1,0){20}}
\put ( 0, 0) {\line (0,1){20}}
\put (10, 0) {\line (0,1){20}}
\put (20,10) {\line (0,1){10}}
\put ( 3, 2) {2}
\put ( 3,12) {1}
\put (13,12) {3}
\end{picture}}
\newcommand{\ya}{
\setlength{\unitlength}{1.0pt}
\begin{picture}(10,30)(0,10)
\thinlines
\put ( 0, 0) {\line (1,0){10}}
\put ( 0,10) {\line (1,0){10}}
\put ( 0,20) {\line (1,0){10}}
\put ( 0,30) {\line (1,0){10}}
\put ( 0, 0) {\line (0,1){30}}
\put (10, 0) {\line (0,1){30}}
\put ( 3, 2) {3}
\put ( 3,12) {2}
\put ( 3,22) {1}
\end{picture}}
\begin{document}
\pagestyle{plain}
\mbox{}\\
\mbox{}\\
\mbox{}\\
\mbox{}\\
\mbox{}\\
\mbox{}\\
\mbox{}\\
\mbox{}\\
\mbox{}\\
\mbox{}\\
\noindent
{\bf ALGEBRAIC TREATMENT OF COLLECTIVE \\
EXCITATIONS IN BARYON SPECTROSCOPY}\\
\mbox{}\\
\mbox{}\\
\mbox{}\\
\indent \hspace{1.9truecm} R.~Bijker$^1$ and A.~Leviatan$^2$\\
\mbox{}\\
\indent \hspace{1.9truecm} $^1$ R.J.~Van de Graaff Laboratory,
University of Utrecht,\\
\indent \hspace{1.9truecm} P.O. Box 80000, 3508 TA Utrecht,
The Netherlands\\
\mbox{}\\
\indent \hspace{1.9truecm} $^2$ Racah Institute of Physics,
The Hebrew University,\\
\indent \hspace{1.9truecm} Jerusalem 91904, Israel\\
\mbox{}\\
\mbox{}\\
\mbox{}\\
{\bf INTRODUCTION}\\
\mbox{}

Algebraic methods have been used extensively in hadronic physics for
the description of the internal degrees of freedom (flavor-spin-color)
\cite{GM}. Spectrum generating algebras and dynamic symmetries have
been very instrumental in the classification of hadronic states and
the construction of mass formulas, such as the Gell-Mann-Okubo
mass formula \cite{Okubo}.

Recently Iachello suggested \cite{FI} to use algebraic methods for
the geometric structure of hadrons as well, and thus, by combining
the degrees of freedom of the internal and relative motion, to obtain
a fully algebraic description of hadrons.
The key ingredient in such an approach is the choice of a spectrum
generating algebra (SGA) for the relevant spatial degrees of freedom,
{\it e.g.} relative quark coordinates or geometric shape variables.
The operators representing physical observables, such as masses,
are then expressed in terms of elements of the algebra.
All matrix elements of interest can be calculated
exactly without having to make further approximations.
As a first application of this approach the string-like configuration
of a quark ($q$) and an antiquark ($\bar q$) in a meson was described
in terms of a $U(4)$ SGA \cite{meson1,meson2}. The $U(4)$ algebra is
realized in terms of bosons whose mutual interactions simulate the
dynamics of the radius vector connecting the quark and antiquark.
This model corresponds to a collective string-like description of
mesons.

The situation for baryons is more complex. In a valence quark
model, baryons are made up of three quarks ($qqq$), some of which
may be identical. Therefore, unlike the situation in mesons,
for baryons one has to construct states with good
permutation symmetry. To fulfill the Pauli principle only particular
combinations of space, spin-flavor and color representations are
allowed. In quark potential models, either nonrelativistic
\cite{IK} or in a relativized form \cite{CI}, baryons are
described as a system of quarks interacting through two-body
interactions. Most calculations are done in a harmonic oscillator basis
and usually only a few oscillator shells are taken into account exactly.
As an alternative to such a single-particle type of description, it is
of interest to consider the possibility of a collective description of
baryons. The quarks participating in a collective motion are
strongly correlated and move coherently. In terms of a harmonic
oscillator basis collectivity requires the mixing of many different
oscillator shells. This is in analogy to the case encountered
in nuclei for which a description of $\alpha$-clustering requires
mixing of many shells in the nuclear shell model.
In any collective description of baryons
one has to mix different oscillator shells in such a way
that the permutation symmetry of the quarks is preserved.
Furthermore, in order to be able to compare the single-particle with
the geometric-collective description of baryons, it would be
beneficial to have a framework that encompasses both points of view.

In this contribution we introduce a $U(7)$ spectrum generating
algebra as a new algebraic string-like model of baryons. This
model satisfies all the above mentioned requirements for a collective
description.
We show explicitly that in the $U(7)$ model it is possible to
incorporate the permutation symmetry of the three quarks exactly
and to accommodate both single-particle and collective types of motion.
The model space consists of many oscillator shells, but since
it is finite dimensional, the $U(7)$ SGA provides a tractable
computational scheme, that involves only the diagonalization
of finite dimensional matrices. A great advantage is that within
the assumptions of the model all calculations can be done
exactly. We discuss some special cases in which we derive closed
analytic solutions that give a clear insight into the
properties of the model.
Finally we note that the same $U(7)$ SGA can be used to describe
rotations and vibrations in triatomic molecules \cite{BDL}.
\mbox{}\\
\mbox{}\\
{\bf THE U(7) QUARK MODEL}\\
\mbox{}

The hamiltonian in quark potential models contains a kinetic
energy term, a confining potential and the hyperfine interaction.
The confining potential is written as the sum of two uncoupled
harmonic oscillators in the Jacobi coordinates and a residual
two-body interaction. The two relative Jacobi coordinates are
\ba
\vec{\rho} &=& (\vec{r}_1 - \vec{r}_2)/\sqrt{2} ~,
\nonumber\\
\vec{\lambda} &=& (\vec{r}_1 + \vec{r}_2 -2\vec{r}_3)/\sqrt{6} ~,
\ea
where $\vec{r}_i$ is the coordinate of the $i$-th quark.
The group structure associated with the harmonic oscillator part
is given by the chain,
\ba U(6) \supset U_{\rho}(3) \otimes U_{\lambda}(3) \supset
SO_{\rho}(3) \otimes SO_{\lambda}(3) \supset SO_{\rho \lambda}(3) ~.
\ea
The unperturbed eigenvalues are given in terms of
$n=n_{\rho}+n_{\lambda}$,
the total number of quanta in the $\rho$ and $\lambda$ oscillators.
The group chain in eq.~(2) provides a convenient basis to diagonalize
the full hamiltonian. Typical calculations involve only a few oscillator
shells. For a given harmonic oscillator shell, $n$, there exist well
known techniques to construct states of good permutation symmetry,
angular momentum and parity \cite{Mosh}.

Collective behavior of quarks inside baryons corresponds to a coherent
motion of the quarks. A description of such correlated motion in a
harmonic oscillator basis requires strong coupling of many different
oscillator shells. In order to accommodate such mixing we
discuss an extension of the $U(6)$ symmetry group of the harmonic
oscillator. From experience gained with algebraic models in nuclear
\cite{ibm} and molecular physics \cite{vibron}, it is known
that this can be achieved by embedding the $U(6)$ algebra in
the compact SGA of $U(7)$. This is done by adding a scalar boson
under the restriction that the total number of bosons is conserved.
The last condition guarantees that the $U(7)$ SGA still describes only
six degrees of freedom, which can be associated with the
two relative Jacobi coordinates.

The building blocks of the $U(7)$ model are the six components of
two dipole bosons with $L^{\pi}=1^-$, denoted in second quantization
by $p_{\rho}^{\dagger}$ and $p_{\lambda}^{\dagger}$, and a scalar boson
with $L^{\pi}=0^+$, denoted by $s^{\dagger}$. A mass operator that
conserves the total number of quanta $N=n_s+n_{\rho}+n_{\lambda}$ can
be expressed in terms of the 49 generators of $U(7)$,
\ba
\begin{array}{ccc}
D_{\lambda} = (p^{\dagger}_{\lambda} s -
s^{\dagger} \tilde{p}_{\lambda})^{(1)} ~, &
A_{\lambda} = i \, (p^{\dagger}_{\lambda} s +
s^{\dagger} \tilde{p}_{\lambda})^{(1)} ~, &
G^{(l)}_{\lambda} = ( p^{\dagger}_{\rho} \tilde{p}_{\rho}
- p^{\dagger}_{\lambda} \tilde{p}_{\lambda} )^{(l)} ~, \\
D_{\rho} = (p^{\dagger}_{\rho} s -
s^{\dagger} \tilde{p}_{\rho})^{(1)} ~, &
A_{\rho} = i \, (p^{\dagger}_{\rho} s +
s^{\dagger} \tilde{p}_{\rho})^{(1)} ~, &
G^{(l)}_{\rho} = ( p^{\dagger}_{\rho} \tilde{p}_{\lambda}
+ p^{\dagger}_{\lambda} \tilde{p}_{\rho} )^{(l)} ~, \\
G^{(l)}_{S} = ( p^{\dagger}_{\rho} \tilde{p}_{\rho}
+ p^{\dagger}_{\lambda} \tilde{p}_{\lambda} )^{(l)} ~, &
G^{(l)}_{A} = i \, ( p^{\dagger}_{\rho} \tilde{p}_{\lambda} -
p^{\dagger}_{\lambda} \tilde{p}_{\rho} )^{(l)} ~, &
\hat{n}_s = s^{\dagger}s ~,
\end{array}
\ea
with $l=0,1,2$ and $\tilde{p}_{m}=(-1)^{1-m} p_{-m}$.
All many-boson states are classified according to the totally symmetric
representation $[N]$ of $U(7)$. The value of $N$ determines the number of
states in the model space. In view of confinement we expect $N$ to be
large. The model space contains several oscillator shells.
For a given value of $N$, it consists of oscillator shells
with $n=n_{\rho}+n_{\lambda}=0,1,\ldots,N$.
In order to construct a mass operator that properly takes into
account the permutation symmetry among the three quarks inside a baryon,
we study the transformation properties of the generators of eq.~(3)
under the permutation group.
\mbox{}\\
\mbox{}\\
{\bf PERMUTATION SYMMETRY IN THE U(7) MODEL}\\
\mbox{}

The harmonic oscillator basis is very well suited to construct
a set of basis states that in addition to good angular momentum
and parity also have good permutation symmetry under the interchange
of any of the three quarks.
For baryons with strangeness $-1$ or $-2$ the only permutation that is
involved is the interchange of the two identical quarks, $P(12)$,
which without loss of generality are taken to be the first two.
For baryons with strangeness 0 or $-3$ the three quarks are
indistinguishable and therefore we have to use in addition to $P(12)$
also the cyclic permutation $P(123)$. All other permutations can be
expressed in terms of these two elementary ones.
Although the $U(7)$ model can describe both strange and nonstrange
baryons, in the present contribution we discuss only
the nonstrange sector, namely the nucleon and the delta
resonances. The spatial wave functions then have to be
combined with a spin-flavor and a color part so that the full
wave function of the baryon is antisymmetric.

In the current approach the spatial part of the wave function is
described in terms of the $U(7)$ SGA. The eigenstates are required
to have well defined transformation properties
under the permutation group, or equivalently, the mass operator for
nonstrange baryon resonances has to be invariant under $S_3$.
The transformation properties under $S_3$ of all operators in the model
follow from those of $s^{\dagger}$, $p^{\dagger}_{\rho}$ and
$p^{\dagger}_{\lambda}$,
\ba
P(12) \left( \begin{array} {l} s^{\dagger} \\ p^{\dagger}_{\rho, m} \\
p^{\dagger}_{\lambda, m} \end{array} \right) &=&
\left( \begin{array}{rrr} 1 & 0 & 0 \\ 0 & -1 & 0 \\ 0 & 0 & 1
\end{array} \right) \left( \begin{array} {l} s^{\dagger} \\
p^{\dagger}_{\rho, m} \\ p^{\dagger}_{\lambda, m} \end{array} \right) ~,
\nonumber\\
\nonumber\\
P(123) \left( \begin{array} {l} s^{\dagger} \\ p^{\dagger}_{\rho, m} \\
p^{\dagger}_{\lambda, m} \end{array} \right) &=&
\left( \begin{array}{rrr} 1 & 0 \qquad & 0 \qquad \\
0 &  \cos 2\pi/3 & \sin 2\pi/3 \\
0 & -\sin 2\pi/3 & \cos 2\pi/3 \end{array} \right)
\left( \begin{array} {l} s^{\dagger} \\ p^{\dagger}_{\rho, m} \\
p^{\dagger}_{\lambda, m} \end{array} \right) ~.
\ea
The $s$-boson is a scalar under the permutation group.
There are three different symmetry classes for the permutation of
three objects: a symmetric one, $S$, an antisymmetric one, $A$, and a
two-dimensional one of mixed symmetry type, denoted by $M_{\rho}$ and
$M_{\lambda}$. The latter have the same transformation properties as
the creation operators, $p^{\dagger}_{\rho}$ and
$p^{\dagger}_{\lambda}$. Alternatively, the three symmetry classes can
be labeled by the irreducible representations of the point group
$D_{3}$ (which is isomorphic to $S_3$) as $A_1$, $A_2$ and $E$,
respectively. It is now straightforward to find the bilinear
combinations of creation and annihilation operators that transform
irreducibly under the permutation group.
The results are shown in table~1.

\begin{table}
\ba
\begin{array}{lcccc}
\hline
& & & & \\
\mbox{Operator} & P(12) & P(123) & S_3 & \mbox{Young tableau} \\
& & & & \\
\hline
& & & & \\
s^{\dagger} ~, \;\; s^{\dagger} s^{\dagger} ~, \;\;
(p^{\dagger}_{\rho} p^{\dagger}_{\rho} +
 p^{\dagger}_{\lambda} p^{\dagger}_{\lambda})^{(l^{\prime})} ~, \;\;
& & & & \\
\hat{n}_s ~, \;\; G^{(l)}_{S} & 1 & 1 & S & \ys \\
& & & & \\
p_{\lambda}^{\dagger} ~, \;\; s^{\dagger} p^{\dagger}_{\lambda} ~, \;\;
(p^{\dagger}_{\rho} p^{\dagger}_{\rho} -
 p^{\dagger}_{\lambda} p^{\dagger}_{\lambda})^{(l^{\prime})} ~, \;\;
& & & & \\
D_{\lambda} ~, \;\; A_{\lambda} ~, \;\; G^{(l)}_{\lambda} &
1 & \lambda & M_{\lambda} & \yml \\
& & & & \\
p_{\rho}^{\dagger} ~, \;\; s^{\dagger} p^{\dagger}_{\rho} ~, \;\;
(p^{\dagger}_{\rho} p^{\dagger}_{\lambda} +
 p^{\dagger}_{\lambda} p^{\dagger}_{\rho})^{(l^{\prime})} ~, \;\;
& & & & \\
D_{\rho} ~, \;\; A_{\rho} ~, \;\; G^{(l)}_{\rho} &
-1 & \rho & M_{\rho} & \ymr \\
& & & & \\
(p^{\dagger}_{\rho} p^{\dagger}_{\lambda})^{(1)} ~, \;\;
G^{(l)}_A & -1 & 1 & A & \ya \\
& & & & \\
\hline
\end{array}
\nonumber
\ea
\caption[]{\small
Transformation properties of linear and bilinear operators
under the $S_3$ permutation group. Here $l=0,1,2$ and
$l^{\prime}=0,2$. \normalsize}
\vspace{10pt}
\end{table}

Next one can use the multiplication rules for $S_3$ to construct
all rotationally invariant interactions that
preserve parity and that are scalars under $S_3$. As a result we find
two terms linear in the $U(7)$ generators of eq.~(3),
\ba
\hat n_s ~, \;\; \hat n = \sqrt{3} \, G^{(0)}_S ~,
\ea
where $\hat n_s = \hat N - \hat n$ is the number operator for scalar
bosons, and $\hat n = \hat n_{\rho} + \hat n_{\lambda}$ counts the
total number of $\rho$ and $\lambda$ bosons. In addition we have the
following quadratic multipole operators,
\ba
& \hat n_s \hat n_s ~, \;\; \hat n_s \hat n ~, \;\;
D_{\rho} \cdot D_{\rho} + D_{\lambda} \cdot D_{\lambda} ~, \;\;
A_{\rho} \cdot A_{\rho} + A_{\lambda} \cdot A_{\lambda} ~,
\nonumber\\
& G^{(l)}_S \cdot G^{(l)}_S ~, \;\;
  G^{(l)}_A \cdot G^{(l)}_A ~, \;\;
  G^{(l)}_{\rho} \cdot G^{(l)}_{\rho} +
  G^{(l)}_{\lambda} \cdot G^{(l)}_{\lambda} ~.
\ea
We note that not all terms in eqs.~(5,6) are independent.
A set of independent two-body interactions that are scalars under the
$S_3$ permutation group can be obtained by transforming to normal
ordered form,
\ba
& s^{\dagger} s^{\dagger} s s ~, \;\;
  s^{\dagger} ( p^{\dagger}_{\rho} \cdot \tilde{p}_{\rho}
+ p^{\dagger}_{\lambda} \cdot \tilde{p}_{\lambda} ) s ~, \;\;
( p^{\dagger}_{\rho} p^{\dagger}_{\lambda} )^{(1)} \cdot
( \tilde p_{\lambda} \tilde p_{\rho} )^{(1)} ~,
\nonumber\\
& ( p^{\dagger}_{\rho} \cdot p^{\dagger}_{\rho}
  + p^{\dagger}_{\lambda} \cdot p^{\dagger}_{\lambda} ) s s
  + s^{\dagger} s^{\dagger} ( \tilde{p}_{\rho} \cdot \tilde{p}_{\rho}
  + \tilde{p}_{\lambda} \cdot \tilde{p}_{\lambda} ) ~,
\nonumber\\
& ( p^{\dagger}_{\rho} p^{\dagger}_{\rho}
  - p^{\dagger}_{\lambda} p^{\dagger}_{\lambda} )^{(l^{\prime})} \cdot
  ( \tilde{p}_{\rho} \tilde{p}_{\rho}
  - \tilde{p}_{\lambda} \tilde{p}_{\lambda} )^{(l^{\prime})}
+ 4 \, ( p^{\dagger}_{\rho} p^{\dagger}_{\lambda})^{(l^{\prime})} \cdot
       ( \tilde p_{\lambda} \tilde p_{\rho})^{(l^{\prime})} ~,
\nonumber\\
& ( p^{\dagger}_{\rho} p^{\dagger}_{\rho}
  + p^{\dagger}_{\lambda} p^{\dagger}_{\lambda} )^{(l^{\prime})} \cdot
  ( \tilde{p}_{\rho} \tilde{p}_{\rho}
  + \tilde{p}_{\lambda} \tilde{p}_{\lambda} )^{(l^{\prime})} ~,
\ea
with $l^{\prime}=0,2$.

Diagonalization of a general $S_3$-invariant mass operator composed
of the terms in eqs.~(5-7) yields
wave functions of good permutation symmetry. The transformation
properties of these wave functions under $S_3$ can be determined
by taking the overlap with harmonic oscillator wave functions that
by construction have good (and known) permutation symmetry
\cite{Mosh}. We used a more direct method which is also based on the
transformation properties under $P(12)$ and $P(123)$.
The symmetry classes ($S,M_{\lambda}$) are distinguished from
($A,M_{\rho}$) by a choice of basis states with the number of
quanta in the $\rho$ oscillator, $n_{\rho}$, even or odd,
respectively. The classes ($S,A$) are distinguished from
($M_{\lambda},M_{\rho}$) by the expectation value $K^2_y$ (with $K_y$
integer) of the operator, $\hat K_y^2=3 \, G^{(0)}_A G^{(0)}_A$,
namely, $|K_y|=0,3,6,\ldots$ for the
former and $|K_y|=1,2,4,5,\ldots$ for the latter. Formally $K_y$
corresponds to the quantum number, $m$, that was used in
ref.~\cite{Bowler} to classify the states in the quark potential
model. It is important to note that our method is valid for any
oscillator shell.
\mbox{}\\
\mbox{}\\
{\bf GEOMETRIC SHAPE AND ITS EXCITATIONS}\\
\mbox{}

The previous analysis shows that already at the level of one- and
two-body interactions, there are many possible terms that can be
included in a $S_3$-invariant mass operator in the $U(7)$ model.
In this section we analyze the model in terms of geometric
variables and study its elementary excitations to gain
a better understanding of the physical content of each of the
allowed interaction terms. This will provide a selection criterion
which terms to include in the mass operator.

A geometric interpretation of the algebraic model can be obtained by
means of a coherent state \cite{Gilmore}, which for the $U(7)$ model
takes the form of a condensate of $N$ bosons,
\ba
\mid N;c \rangle &=&
\frac{1}{\sqrt{N!}} \Bigl( b_c^{\dagger} \Bigr)^N \mid 0 \rangle ~,
\ea
with
\ba
b_c^{\dagger} &=& (1+R^2)^{-1/2} \Bigl[ s^{\dagger}
+ r_{\rho} \, p_{\rho,0 }^{\dagger} + r_{\lambda} \sum_{m}
d^{(1)}_{m, 0}(\theta) \, p_{\lambda, m}^{\dagger} \Bigr] ~.
\ea
Here $R=\sqrt{r_{\rho}^2+r_{\lambda}^2}$. The two vectors,
$\vec{r}_{\rho}$ and $\vec{r}_{\lambda}$, in the condensate span the
$xz$-plane. We have chosen the $z$-axis along the direction of
$\vec{r}_{\rho}=r_{\rho} \hat{z}$, and $\vec{r}_{\lambda}$ is rotated
by an angle $\theta$ about the out-of-plane $y$-axis,
$\vec{r}_{\rho}\cdot\vec{r}_{\lambda}=r_{\rho}r_{\lambda}\cos\theta$.

The expectation value of the mass operator in the condensate
defines a classical potential function, $V(r_{\rho}r_{\lambda}\theta)=
\langle N;c \mid \hat M^2 \mid N;c \rangle$.
The equilibrium shape is determined by minimizing the potential
function with respect to the coordinates, $r_{\rho}$ and
$r_{\lambda}$, and the relative angle $\theta$.
The equilibrium values of these shape parameters are denoted by
$\bar{r}_{\rho}$, $\bar{r}_{\lambda}$ and $\bar{\theta}$,
respectively. For the problem at hand, three identical quarks
inside a baryon, the most general $S_3$-invariant $U(7)$ mass operator
with one- and two-body terms yields a rigid nonlinear equilibrium
shape characterized by
\ba
\bar{r}_{\rho}=\bar{r}_{\lambda}~, \hspace{1cm}
\bar{\theta}=\frac{\pi}{2} ~.
\ea
These are precisely the conditions satisfied by the Jacobi coordinates
of eq.~(1) for an equilateral triangular shape. This strongly suggests to
associate these coordinates with the algebraic shape parameters in
eq.~(9). The nonnegative equilibrium value
$\bar{r}_{\rho}=\bar{r}_{\lambda}=\bar{r}$ is
a measure of the dipole deformation in the ground state condensate.
For the special case of $\bar{r}=0$ we recover the spherical
condensate of the quark potential model.

The equilibrium configuration of three identical quarks in a
baryon is represented in the $U(7)$ model by an intrinsic state,
eq.~(8), composed of $N$ condensate bosons with
$r_{\rho}=r_{\lambda}$ and $\theta=\pi/2$:
\ba
b_c^{\dagger} &=& (1+R^2)^{-1/2} \Bigl\{ s^{\dagger}
+ R \frac{1}{\sqrt{2}} \Bigl[ p_{\rho,0 }^{\dagger} - \frac{1}{\sqrt{2}}
( p_{\lambda, 1}^{\dagger} - p_{\lambda,-1}^{\dagger})
\Bigr] \Bigr\} ~.
\ea
It corresponds geometrically to an equilateral triangular shape with
the $y$-axis as a threefold symmetry axis.
Excitations of the equilibrium shape are represented
by the following six deformed bosons which, together with
$b^{\dagger}_c$, form a complete orthonormal basis,
\ba
b^{\dagger}_{u} &=& (1+R^{2})^{-1/2} \Bigl\{ - R \, s^{\dagger}
+ \frac{1}{\sqrt{2}} \Bigl[ p_{\rho,0 }^{\dagger} - \frac{1}{\sqrt{2}}
\Bigl( p_{\lambda, 1}^{\dagger} - p_{\lambda,-1}^{\dagger} \Bigr)
\Bigr] \Bigr\} ~,
\nonumber\\
b^{\dagger}_{v} &=& \frac{1}{\sqrt{2}}
\Bigl[ p_{\rho,0 }^{\dagger} + \frac{1}{\sqrt{2}}
\Bigl( p_{\lambda, 1}^{\dagger} - p_{\lambda,-1}^{\dagger} \Bigr)
\Bigr] ~,
\nonumber\\
b^{\dagger}_{w} &=& \frac{1}{\sqrt{2}} \Bigl[ - \frac{1}{\sqrt{2}}
\Bigl( p_{\rho, 1}^{\dagger} - p_{\rho,-1}^{\dagger} \Bigr)
+ p^{\dagger}_{\lambda, 0} \Bigr] ~,
\nonumber\\
b^{\dagger}_{x} &=& \frac{1}{\sqrt{2}}
\Bigl( p^{\dagger}_{\rho,1} + p^{\dagger}_{\rho,-1} \Bigr) ~,
\nonumber\\
b^{\dagger}_{y} &=& \frac{1}{\sqrt{2}} \Bigl[ \frac{1}{\sqrt{2}}
\Bigl( p^{\dagger}_{\rho,1} - p^{\dagger}_{\rho,-1} \Bigr)
+ p_{\lambda, 0}^{\dagger} \Bigr] ~,
\nonumber\\
b^{\dagger}_{z} &=& \frac{1}{\sqrt{2}}
\Bigl( p_{\lambda, 1}^{\dagger} + p_{\lambda,-1}^{\dagger} \Bigr) ~.
\ea
The above boson operators represent excitations of the condensate
which involve three vibrational intrinsic modes and three rotational
Goldstone modes.
The vibrational modes are a symmetric radial mode ($b^{\dagger}_{u}$),
an antisymmetric radial mode ($b^{\dagger}_{v}$) and an angular mode
($b^{\dagger}_{w}$). The boson operators $b^{\dagger}_{i}$ $(i=x,y,z)$
are associated with rotational modes. They are obtained by applying
the three components of the angular momentum operator, $\hat L_{i}$
($i=x,y,z$), on the condensate (8) with the shape parameters of
eq.~(10). The physical interpretation of these excitation modes are
discussed in more detail in the following sections.

The vibrational and rotational excitations of a $S_3$-invariant
$U(7)$ mass operator can be studied by decomposing it into
an intrinsic (vibrational) and a collective (rotational) part,
\ba
\hat M^2_{U(7)} &=& \hat M^{2}_{\mbox{int}}
+ \hat M^{2}_{\mbox{coll}} ~.
\ea
This resolution is obtained \cite{LK} by requiring that
the intrinsic part annihilates the ground state condensate and has the
same shape of the potential function as the original mass operator.
The collective part of the mass operator has by construction
a completely flat potential function.
We now discuss each of these parts separately.
\newpage
\noindent
{\bf VIBRATIONS}\\
\mbox{}

\begin{figure}
\setlength{\unitlength}{1pt}
\begin{picture}(240,200)(-100,0)
\thinlines
\put (  0,  0) {\line(1,0){240}}
\put (  0,  0) {\line(0,1){200}}
\put (  0,200) {\line(1,0){240}}
\put (240,  0) {\line(0,1){200}}
\put (  0, 50) {\line(1,0){5}}
\put (  0,100) {\line(1,0){5}}
\put (  0,150) {\line(1,0){5}}
\put (-75,175) {$M^2$ (GeV$^2$)}
\put (-20, 45) {1}
\put (-20, 95) {2}
\put (-20,145) {3}
\multiput (80,44)(5,0){29}{\circle*{0.1}}
\thicklines
\put ( 20, 44) {\line(1,0){60}}
\put ( 90,104) {\line(1,0){60}}
\put (160,146) {\line(1,0){60}}
\thinlines
\put ( 20, 54) {$N$(939)$P_{11}$}
\put ( 90,114) {$N$(1440)$P_{11}$}
\put (160,156) {$N$(1710)$P_{11}$}
\put ( 45, 25) {$A_1$}
\put (115, 25) {$A_1$}
\put (185, 25) {$E$}
\put (120, 44) {\vector(0, 1){ 60}}
\put (120,104) {\vector(0,-1){ 60}}
\put (190, 44) {\vector(0, 1){102}}
\put (190,146) {\vector(0,-1){102}}
\end{picture}
\caption[]{\small Schematic representation of the vibrational spectrum
of nucleon resonances. The resonances are labeled by the usual
spectroscopic notation \cite{PDG} and their vibrational permutation
symmetry.
\normalsize}
\vspace{14.5pt}
\end{figure}
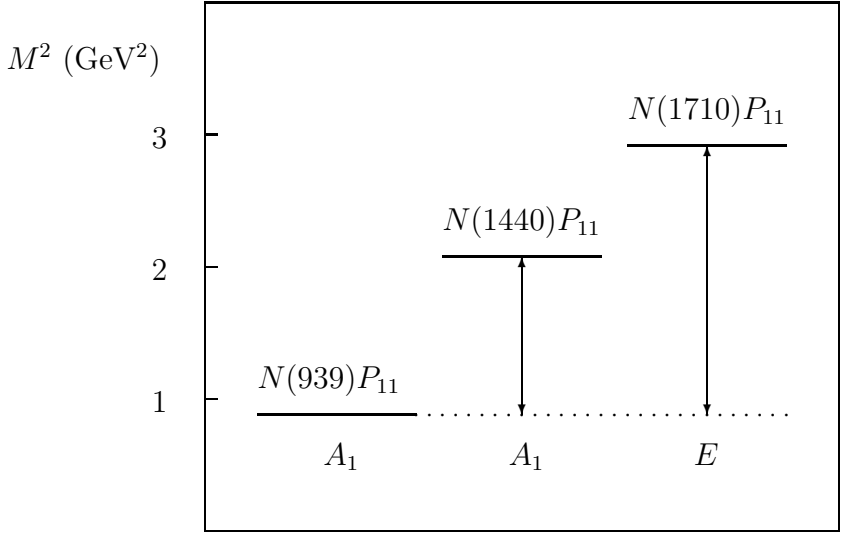

The intrinsic part of a $S_3$-invariant $U(7)$ mass operator
that annihilates the condensate with $r_{\rho}=r_{\lambda}=\bar{r} \neq 0$
and $\theta=\bar{\theta}=\pi/2$, consists of two terms only,
\ba
\hat M^{2}_{\mbox{int}} &=& \xi_1 \,
\Bigl ( R^2 \, s^{\dagger} s^{\dagger}
- p^{\dagger}_{\rho} \cdot p^{\dagger}_{\rho}
- p^{\dagger}_{\lambda} \cdot p^{\dagger}_{\lambda} \Bigr ) \,
\Bigl ( R^2 \, s s - \tilde{p}_{\rho} \cdot \tilde{p}_{\rho}
- \tilde{p}_{\lambda} \cdot \tilde{p}_{\lambda} \Bigr )
\nonumber\\
&& + \xi_2 \, \Bigl [
\Bigl ( p^{\dagger}_{\rho} \cdot p^{\dagger}_{\rho}
- p^{\dagger}_{\lambda} \cdot p^{\dagger}_{\lambda} \Bigr ) \,
\Bigl ( \tilde{p}_{\rho} \cdot \tilde{p}_{\rho}
- \tilde{p}_{\lambda} \cdot \tilde{p}_{\lambda} \Bigr )
+ 4 \, \Bigl ( p^{\dagger}_{\rho} \cdot p^{\dagger}_{\lambda} \Bigr ) \,
\Bigl ( \tilde p_{\lambda} \cdot \tilde p_{\rho} \Bigr ) \Bigr ] ~,
\ea
with $R^2 = 2\bar{r}^2$.
The vibrational excitations can be obtained from a normal mode analysis
of the intrinsic part of the mass operator.
This is done \cite{LK} by rewriting the
mass operator in terms of the deformed bosons of eqs.~(11,12),
replacing the condensate bosons, $b_c$ and $b_c^{\dagger}$, by their
classical mean field value, $\sqrt{N}$, and keeping terms of
leading order in $N$. In the large $N$ limit the normal modes are
diagonal in the deformed boson basis and the mass operator reduces to
a harmonic oscillator form,
\ba
\frac{1}{N} \hat M^{2}_{\mbox{int}} = \lambda_1 \, b_u^{\dagger} b_u
+ \lambda_2 \, \Bigl ( b_v^{\dagger} b_v + b_w^{\dagger} b_w \Bigr )
+ {\cal O}(1/\sqrt{N}) ~,
\ea
with eigenvalues
\ba
\lambda_1 &=& 4 \, \xi_1 R^2 ~,
\nonumber\\
\lambda_2 &=& 4 \, \xi_2 R^2 / (1+R^2) ~.
\ea
Eq.~(15) identifies the deformed bosons that correspond to the
three fundamental vibrations: a symmetric stretching ($u$),
an antisymmetric stretching ($v$) and a bending vibration ($w$).
The first two are radial excitations, whereas the third is an angular
mode which corresponds to oscillations in the angle $\theta$ between
the two Jacobi coordinates. The angular mode is degenerate with the
antisymmetric radial mode (with eigenvalue $\lambda_2$). This is
in agreement with the point group classification of the fundamental
vibrations for a symmetric X$_3$ configuration \cite{Herzberg}.
This shows that $\hat M^{2}_{\mbox{int}}$
describes the vibrational excitations of an oblate symmetric top.
The rotational bosons, $b^{\dagger}_i$ $(i=x,y,z)$, are missing from
eq.~(15) and correspond to massless Goldstone bosons associated
with the rotation symmetry which is spontaneously broken in the
condensate.

The intrinsic part of the mass operator describes the vibrational
excitations of baryon resonances. In the large $N$ limit the
vibrational spectrum is harmonic,
\ba
M^{2}_{\mbox{vib}} = N \left[ \lambda_1 \, n_u
+ \lambda_2 \, (n_v + n_w) \right] ~.
\ea
In the nucleon sector, the vibrationless ground state
($n_u=n_v=n_w=0$) is identified with the nucleon $N$(939).
The symmetric stretching vibration with frequency, $\lambda_1$, has
the same symmetry properties as the ground state. It is therefore
possible to associate the vibration ($n_u=1, n_v+n_w=0$)
with the Roper resonance $N$(1440). The $N$(1710) resonance could be
a candidate for the other (two-dimensional)
vibrational mode with ($n_u=0, n_v+n_w=1$).
Similar considerations apply to the delta sector.

A great advantage of this analysis is that we have established a
relation between the parameters of otherwise abstract `algebraic'
interactions, and measured experimental quantities, in this case
the mass of specific resonances. The vibrational contribution to
the masses, eq.~(17), is obtained in the large $N$ approximation which
is expected to be valid in view of confinement. The extracted value
of the parameters, $\xi_1$ and $\xi_2$, can be used as a good starting
value in a more detailed fitting procedure in an exact numerical
calculation. In figure~1 we show a schematic representation of
the vibrational spectrum of nonstrange baryon resonances.
\mbox{}\\
\mbox{}\\
{\bf ROTATIONS}\\
\mbox{}

On top of each vibrational excitation there is a whole series of
rotational states. In a geometric description the rotational
excitations are labeled by the
angular momentum, $L$, its projection on the threefold symmetry
axis, $K=0,1,\ldots,$ parity, and the transformation character
under the permutation group. For a given value of $K$, the states
have angular momentum $L=K,K+1,\ldots,$ and parity $\pi=(-)^K$.
For rotational states built on vibrations of type $A_1$
(symmetric under $S_3$) each $L$ state is single for $K=0$ and
twofold degenerate for $K \neq 0$. For rotational states built on
vibrations of type $E$ (mixed $S_3$ symmetry) each $L$ state is
twofold degenerate for $K=0$ and fourfold degenerate for $K\neq 0$.
The transformation property of the states under the permutation group
is found by multiplying the symmetry character of the vibrational and
the rotational wave functions. The results are shown schematically
in figure~2.

\begin{figure}
\setlength{\unitlength}{1pt}
\begin{picture}(440,210)(0,-10)
\thinlines
\put (  0,-10) {\line(1,0){440}}
\put (  0,200) {\line(1,0){440}}
\put (  0,-10) {\line(0,1){210}}
\put (205,-10) {\line(0,1){210}}
\put (440,-10) {\line(0,1){210}}
\thicklines
\put (10, 50) {\line(1,0){25}}
\put (10, 70) {\line(1,0){25}}
\put (10,110) {\line(1,0){25}}
\put (10,170) {\line(1,0){25}}
\put (55, 66) {\line(1,0){25}}
\put (55,106) {\line(1,0){25}}
\put (55,166) {\line(1,0){25}}
\put (100, 94) {\line(1,0){25}}
\put (100,154) {\line(1,0){25}}
\put (145,134) {\line(1,0){25}}
\thinlines
\put ( 60,  5) {$A_1$-type vibration}
\put ( 10, 25) {$K$=0}
\put ( 55, 25) {$K$=1}
\put (100, 25) {$K$=2}
\put (145, 25) {$K$=3}
\put ( 37, 50) {$0^+_{A_1}$}
\put ( 37, 70) {$1^+_{A_2}$}
\put ( 37,110) {$2^+_{A_1}$}
\put ( 37,170) {$3^+_{A_2}$}
\put ( 82, 66) {$1^-_{E}$}
\put ( 82,106) {$2^-_{E}$}
\put ( 82,166) {$3^-_{E}$}
\put (127, 94) {$2^+_{E}$}
\put (127,154) {$3^+_{E}$}
\put (172,134) {$3^-_{A_1 A_2}$}
\thicklines
\put (215, 50) {\line(1,0){25}}
\put (215, 70) {\line(1,0){25}}
\put (215,110) {\line(1,0){25}}
\put (215,170) {\line(1,0){25}}
\put (260, 66) {\line(1,0){25}}
\put (260,106) {\line(1,0){25}}
\put (260,166) {\line(1,0){25}}
\put (325, 94) {\line(1,0){25}}
\put (325,154) {\line(1,0){25}}
\put (390,134) {\line(1,0){25}}
\thinlines
\put (280,  5) {$E$-type vibration}
\put (215, 25) {$K$=0}
\put (260, 25) {$K$=1}
\put (325, 25) {$K$=2}
\put (390, 25) {$K$=3}
\put (242, 50) {$0^+_E$}
\put (242, 70) {$1^+_E$}
\put (242,110) {$2^+_E$}
\put (242,170) {$3^+_E$}
\put (287, 66) {$1^-_{A_1 A_2 E}$}
\put (287,106) {$2^-_{A_1 A_2 E}$}
\put (287,166) {$3^-_{A_1 A_2 E}$}
\put (352, 94) {$2^+_{A_1 A_2 E}$}
\put (352,154) {$3^+_{A_1 A_2 E}$}
\put (417,134) {$3^-_{E E}$}
\end{picture}
\caption[]{\small
Schematic representation of the rotational structure built on
vibrations of type $A_1$ (lefthand side) and vibrations of type $E$
(righthand side). The levels are labeled by $K$, $L^{\pi}_t$, where
$t$ denotes the overall (vibrational plus rotational) permutation
symmetry. Each $E$ state is doubly degenerate.
\normalsize}
\vspace{14.5pt}
\end{figure}
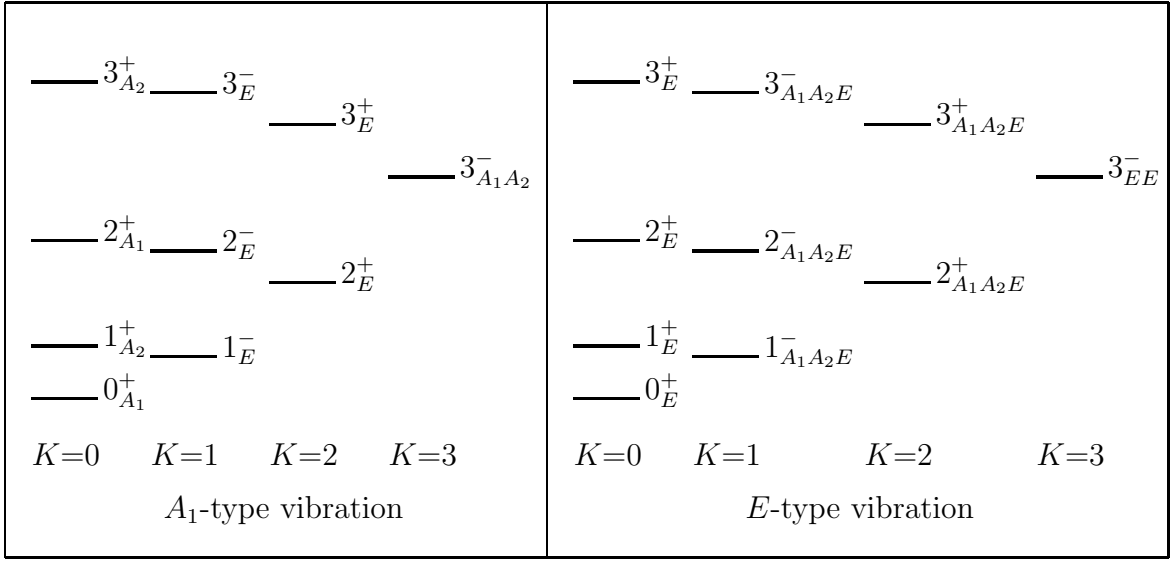

The collective part of the full mass operator contains the
contribution of rotations to the mass. By construction
$\hat M^{2}_{\mbox{coll}}$ consists of interaction terms which do not
affect the shape of the potential function.
Apart from the one-body term, $\hat n$, whose contribution to the mass
is negligible for large $N$, the collective part of the mass
operator can be expressed in terms of the Casimir operators of the
following group chain,
\ba
U(7) \supset SO(7) \supset SO(6) \supset SO(3) \otimes SO(2) ~.
\ea
The bosonic Casimir invariants of $U(7)$ involve only the total
number operator, $\hat N= \hat n_s + \hat n$, which is a
conserved quantity. The only terms in the collective part of the
mass operator that contribute to the excitation spectrum are
therefore,
\ba
\hat M^{2}_{\mbox{coll}} = \kappa_{1} \, \hat C_{SO(7)}
+ \kappa_{2} \, \hat C_{SO(6)}
+ \kappa_{3} \, \hat C_{SO(3)} + \kappa_{4} \, [\hat C_{SO(2)}]^2 ~.
\ea
Explicit expressions for the Casimir operators ($\hat C_G$) in terms of
the generators of the groups ($G$) are shown in table~2.

The last two terms in the collective part of the mass operator (19)
commute with any $S_3$-invariant mass operator and thus correspond to
{\em exact} symmetries. Their eigenvalues are similar in form to those
of a symmetric top, $\kappa_{3} \, L(L+1) + \kappa_{4} \, K^2_y$.
For the lowest eigenstates of each $L$ the value of $K_y$ coincides
with that of the projection of the angular momentum on the threefold
symmetry axis, denoted previously by $K$.
Since $L$ and $K_y$ are always good quantum numbers, it is
straightforward to include in the collective part of the mass operator
higher orders of the corresponding Casimir operators.
Generalizing the mass operator to contain arbitrary functions,
$f(\hat L \cdot \hat L)$ and $g(\hat K^2_y)$, the expression for the
resulting rotational spectrum becomes $f[L(L+1)] + g[K^2_y]$.

\begin{table}
\ba
\begin{array}{cccc}
\hline
& & & \\
\mbox{Group } G & \mbox{Generators} & &
\mbox{Casimir operator } \hat C_G \\
& & & \\
\hline
& & & \\
SO(7) & G^{(1)}_{S}, G^{(1)}_{\rho}, G^{(1)}_{\lambda},
G^{(0)}_{A}, G^{(2)}_{A}, A_{\rho}, A_{\lambda} & & \hat C_{SO(6)}
+ A_{\rho} \cdot A_{\rho} + A_{\lambda} \cdot A_{\lambda} \\
& & & \\
SO(6) & G^{(1)}_{S}, G^{(1)}_{\rho}, G^{(1)}_{\lambda},
G^{(0)}_{A}, G^{(2)}_{A} & & G^{(1)}_{S} \cdot G^{(1)}_{S}
+ G^{(1)}_{\rho} \cdot G^{(1)}_{\rho}
+ G^{(1)}_{\lambda} \cdot G^{(1)}_{\lambda} \\
& & &  + G^{(0)}_{A} \cdot G^{(0)}_{A}
       + G^{(2)}_{A} \cdot G^{(2)}_{A} \\
& & & \\
SO(3) & \hat L  = \sqrt{2} \, G^{(1)}_{S} & & \hat L \cdot \hat L \\
& & & \\
SO(2) & \hat K_y  = -\sqrt{3} \, G^{(0)}_{A} & & \hat K_y \\
& & & \\
\hline
\end{array}
\nonumber
\ea
\caption[]{\small
Generators and Casimir operators of the groups relevant for
the collective part \\ of the mass operator.
\normalsize}
\vspace{10pt}
\end{table}

In general the first two terms in eq.~(19) do not correspond to exact
symmetries of a $S_3$-invariant mass operator and do not commute
with the intrinsic part of the mass operator, eq.~(14).
Therefore, in addition to shifting and splitting the bands generated
by $\hat M^2_{\mbox{int}}$, they can also mix them. Their effect on
the spectrum can be studied numerically. However, just as was done
for the intrinsic part of the mass operator, we can gain physical
insight of the different rotational terms in the collective part of
the mass operator, by studying the large $N$ limit,
\ba
\frac{1}{N} \hat M^{2}_{\mbox{coll}} &=&
- \eta_1 \, (b_u^{\dagger}-b_u)^2
- \eta_2 \, \Bigl [ (b_v^{\dagger}-b_v)^2
                  + (b_w^{\dagger}-b_w)^2 \Bigr ]
\nonumber\\
&& + \eta_3 \, \Bigl [ (b_x^{\dagger}+b_x)^2
                     + (b_z^{\dagger}+b_z)^2 \Bigr ]
   - \eta_4 \, (b_y^{\dagger}-b_y)^2 + {\cal O}(1/\sqrt{N}) ~,
\ea
with
\ba
\eta_1 &=& \kappa_1 ~,
\nonumber\\
\eta_2 &=& (1+R^2)^{-1} [ \kappa_1 + \kappa_2 R^2 ] ~,
\nonumber\\
\eta_3 &=& (1+R^2)^{-1}
[ \kappa_1 + (\kappa_2 + \tfrac{1}{2}\kappa_3) R^2 ] ~,
\nonumber\\
\eta_4 &=& (1+R^2)^{-1}
[ \kappa_1 + (\kappa_2+\kappa_3 +\kappa_4) R^2 ] ~.
\ea
The $b_x$, $b_y$ and $b_z$ bosons are the Goldstone bosons connected
with rotations in configuration space, whereas the $b_u$, $b_v$
and $b_w$ bosons correspond to generalized rotations in a higher
dimensional space. It is seen from eqs.~(20,21) that the
$\kappa_1$-term
of eq.~(19) is associated with rotations in a six-dimensional space
($u,v,w,x,y,z$) space and the $\kappa_2$-term with rotations in a
five-dimensional subspace ($v,w,x,y,z$). The $\kappa_3$-term,
$\hat L \cdot \hat L$, corresponds to ordinary three-dimensional
($x,y,z$) rotations and the $\kappa_4$-term, $\hat K^2_y$, is associated
with a rotation about the threefold symmetry $y$-axis.
The equality of the coefficients of the ($v,w$) and the ($x,z$) terms
in eq.~(20) is a reflection of the oblate-top nature of the
$S_3$-invariant mass operator.

Summarizing, we have shown that a $S_3$-invariant $U(7)$ mass operator
corresponds geometrically to rotations and vibrations of an equilateral
equilibrium configuration of three quarks in a baryon.
There are only six independent terms
in the mass operator that determine the excitation spectrum: two for
the vibrational excitations and four for the rotational excitations.
The two vibrational terms cluster the states into bands.
Two of the rotational terms correspond to exact symmetries of the mass
operator and their eigenvalues are obtained in closed form.
Whereas these terms are diagonal and hence only cause band splitting,
the remaining two rotational terms may in addition cause band mixing.
\mbox{}\\
\mbox{}\\
{\bf BARYON RESONANCES}\\
\mbox{}

In the previous sections we introduced a $U(7)$ spectrum generating
algebra for the spatial part of the baryon wave function.
The total wave function for baryon resonances consists of a spatial,
a spin-flavor and a color part,
\ba
\psi = \psi_L \, \phi_{sf} \, \psi_c ~.
\ea
To satisfy the Pauli principle for a system of identical quarks
these parts have to be combined so that the total wave function
is antisymmetric.

Both the spatial and the spin-flavor degrees of freedom contribute to
the mass operator. In principle there could be also spin-flavor
dependence in the coefficients of the interaction terms in
$\hat M^2_{\mbox{space}}$. Although these effects can be included
without difficulty in an algebraic approach, in the present study we
do not take them into account and write
\ba
\hat M^2 = \hat M^{2}_{\mbox{space}}
         + \hat M^{2}_{\mbox{spin-flavor}} ~.
\ea

For the spin-flavor part we consider the
$SU(6) \supset  SU(3) \otimes SU(2)$ dynamic symmetry of G\"ursey and
Radicati \cite{GR}. In general, this symmetry may be broken (as is the
case with the hyperfine interaction in the quark potential model), which
would lead to coupling terms in eq.~(23). Here, for simplicity,
we limit ourselves to a diagonal breaking for
which the eigenvalues are given in closed form,
\ba
M^2_{\mbox{spin-flavor}}
= a \, \Bigl [ \langle \hat C_{SU(6)} \rangle - 45 \Bigr ]
+ b \, \Bigl [ \langle \hat C_{SU(3)} \rangle -  9 \Bigr ]
+ c \, S(S+1) ~.
\ea
The first term involves the Casimir operator of the $SU(6)$
spin-flavor group with eigenvalues 45, 33 and 21 for the
representations $56 \leftrightarrow A_1$, $70 \leftrightarrow E$ and
$20 \leftrightarrow A_2$, respectively. The second term
involves the Casimir invariant of the $SU(3)$ flavor group with
eigenvalues 9 and 18 for the octet and decuplet, respectively.
The last term contains the eigenvalues $S(S+1)$ of the total spin
operator.

For the spatial part of the mass operator we take the $S_3$
invariant $U(7)$ mass operator discussed in the previous section
which is decomposed into an intrinsic and a collective part
\ba
\hat M^2_{\mbox{space}} = \hat M^2_{U(7)} =
\hat M^{2}_{\mbox{int}} + \hat M^{2}_{\mbox{coll}} ~.
\ea
Here, for simplicity, we take a simplified form for the collective
part of the mass operator, containing only spatial rotations
\ba
\hat M^2_{\mbox{coll}} \rightarrow \hat M^{2}_{\mbox{rot}}
= \alpha \, \sqrt{\hat L \cdot \hat L + \frac{1}{4}} ~,
\ea
with eigenvalues
\ba
M^{2}_{\mbox{rot}} = \alpha \, (L+1/2) ~.
\ea

The total mass operator can now be diagonalized numerically to get a
fit for the masses of nonstrange baryon resonances. Instead, for the
vibrational part we use the large $N$ expression in eq.~(17) with
$R^2=1$, and obtain the following analytic expression for the masses,
\ba
M^2 = M_{0}^{2} + M^{2}_{\mbox{vib}}
+ M^{2}_{\mbox{rot}} + M^{2}_{\mbox{spin-flavor}} ~.
\ea
Here $M_{0}^{2}$ is a constant and the other contributions are given
by eqs.~(17,24,27).

For the nucleon which is a member of the flavor octet the relevant mass
formulas read
\ba
M^2_N(A_1,L,S=1/2) - M^2_{N(939)} &=& M^2_{\mbox{vib}}
+ \alpha \, L ~,
\nonumber \\
M^2_N(E,L,S=1/2) - M^2_{N(939)} &=& M^2_{\mbox{vib}}
+ \alpha \, L - 12a ~,
\nonumber \\
M^2_N(E,L,S=3/2) - M^2_{N(939)} &=& M^2_{\mbox{vib}}
+ \alpha \, L - 12a + 3c ~,
\nonumber \\
M^2_N(A_2,L,S=1/2) - M^2_{N(939)} &=& M^2_{\mbox{vib}}
+ \alpha \, L - 24a ~.
\ea
Here $M^2_{N(939)}=0.882 \mbox{ GeV}^2=M^2_0 + \alpha/2 + 3c/4$.
For the delta which is a member of the flavor decuplet the relevant
mass formulas read
\ba
M^2_{\Delta}(A_1,L,S=3/2) - M^2_{\Delta(1232)} &=& M^2_{\mbox{vib}}
+ \alpha \, L ~,
\nonumber \\
M^2_{\Delta}(E,L,S=1/2) - M^2_{\Delta(1232)} &=& M^2_{\mbox{vib}}
+ \alpha \, L - 12a - 3c ~,
\ea
\newpage
\begin{center}
\begin{tabular}{cccccccccc}
\hline
 & & & & & & & & & \\
Mass & Status & $M^{2}_{\mbox{exp}}$ & $J^{\pi}$ & $L^{\pi},K$ & $S$
& $t$ & $M^{2}_{\mbox{calc}}$ & $\%$ Error & $(D,L^{\pi}_{n})S$ \\
& & & & & & & & & \\
\hline
& & & & & & & & & \\
$N(939)P_{11}$ & **** & 0.882 & ${1\over 2}^{+}$ & $0^{+},0$ &
${1\over 2}$ & $A_{1}$ & 0.882 & 0 & $(56,0^{+}_{0}){1\over 2}$ \\
& & & & & & & & & \\
$N(1440)P_{11}$ & **** & 2.074 & ${1\over 2}^{+}$ & $0^{+},0$ &
${1\over 2}$ & $A_{1}$ & 2.074 & 0 & $(56,0^{+}_{2}){1\over 2}$ \\
& & & & & & & & & \\
$N(1520)D_{13}$ & **** & 2.310 & ${3\over 2}^{-}$ & $1^{-},1$ &
${1\over 2}$ & $E$ & 2.442 &$-$5.7 & $(70,1^{-}_{1}){1\over 2}$ \\
& & & & & & & & & \\
$N(1535)S_{11}$ & **** & 2.356 & ${1\over 2}^{-}$ & $1^{-},1$ &
${1\over 2}$ & $E$ & 2.442 &$-$3.6 & $(70,1^{-}_{1}){1\over 2}$ \\
& & & & & & & & & \\
$N(1650)S_{11}$ & **** & 2.722 & ${1\over 2}^{-}$ & $1^{-},1$ &
${3\over 2}$ & $E$ & 2.817 &$-$3.5 & $(70,1^{-}_{1}){3\over 2}$ \\
& & & & & & & & & \\
$N(1675)D_{15}$ & **** & 2.806 & ${5\over 2}^{-}$ & $1^{-},1$ &
${3\over 2}$ & $E$ & 2.817 &$-$0.4 & $(70,1^{-}_{1}){3\over 2}$ \\
& & & & & & & & & \\
$N(1680)F_{15}$ & **** & 2.822 & ${5\over 2}^{+}$ & $2^{+},0$ &
${1\over 2}$ & $A_{1}$ & 2.994 &$-$6.0 & $(56,2^{+}_{2}){1\over 2}$ \\
& & & & & & & & & \\
$N(1700)D_{13}$ & *** & 2.890 & ${3\over 2}^{-}$ & $1^{-},1$ &
${3\over 2}$ & $E$ & 2.817 & 2.5 & $(70,1^{-}_{1}){3\over 2}$ \\
& & & & & & & & & \\
$N(1710)P_{11}$ & *** & 2.924 & ${1\over 2}^{+}$ & $0^{+},0$ &
${1\over 2}$ & $E$ & 2.924 & 0 & $(70,0^{+}_{2}){1\over 2}$ \\
& & & & & & & & & \\
$N(1720)P_{13}$ & **** & 2.958 & ${3\over 2}^{+}$ & $2^{+},0$ &
${1\over 2}$ & $A_{1}$ & 2.994 &$-$1.2 & $(56,2^{+}_{2}){1\over 2}$ \\
& & & & & & & & & \\
$N(2190)G_{17}$ & **** & 4.796 & ${7\over 2}^{-}$ & $3^{-},1$ &
${1\over 2}$ & $E$ & 4.554 & 5.0 & $(70,3^{-}_{3}){1\over 2}$ \\
& & & & & & & & & \\
                &      &       &                  &           &
${3\over 2}$ & $E$ & 4.929 &$-$2.8 &                            \\
& & & & & & & & & \\
$N(2220)H_{19}$ & **** & 4.928 & ${9\over 2}^{+}$ & $4^{+},0$ &
${1\over 2}$ & $A_{1}$ & 5.106 &$-$3.6 & $(56,4^{+}_{4}){1\over 2}$ \\
& & & & & & & & & \\
$N(2250)G_{19}$ & **** & 5.063 & ${9\over 2}^{-}$ & $3^{-},1$ &
${3\over 2}$ & $E$ & 4.929 & 2.6 & $(70,3^{-}_{3}){3\over 2}$ \\
& & & & & & & & & \\
$N(2600)I_{1,11}$ & *** & 6.760 & ${11\over 2}^{-}$ & $5^{-},1/5$
& ${1\over 2}$ & $E$ & 6.666 & 1.4 & \\
& & & & & & & & & \\
\hline
\end{tabular}
\end{center}
\noindent
Table~3: {\small Oblate top classification of nonstrange baryons
of the $N$ family with $I=1/2$. Here $t$ denotes the overall permutation
symmetry. The last column lists the dominant representation of the quark
model assignment in a $SU_{sf}(6) \otimes O(3)$ basis \cite{PDG}.
$M^2$ is given in GeV$^2$.
The experimental values are taken from \cite{PDG}.
\normalsize}\\
\vspace{14.5pt}\\
\noindent
and $M^2_{\Delta(1232)}=1.518 \mbox{ GeV}^2=M^2_{N(939)} + 9b + 3c$.
The parameters $M^2_0$, $b$, $\xi_1 N$ and $\xi_2 N$ are determined
from the mass squared of $N(939)$, $\Delta(1232)$, $N(1440)$ and
$N(1710)$ respectively. The remaining parameters $a$, $c$, and
$\alpha$ are determined by fitting the mass of the baryon resonances
with *** or **** status. The values of the parameters (in GeV$^2$)
extracted in a least square fit are
\ba
& M_{0}^{2}=0.260 ~, \;\; \xi_1 N=0.298 ~, \;\;
\xi_2 N=0.769 ~, \;\; \alpha=1.056 ~,
\nonumber\\
& a=-0.042 ~, \;\; b=0.029 ~, \;\; c=0.125 ~.
\ea
\newpage
\begin{center}
\begin{tabular}{cccccccccc}
\hline
 & & & & & & & & & \\
Mass & Status & $M^{2}_{\mbox{exp}}$ & $J^{\pi}$ & $L^{\pi},K$ & $S$
& $t$ & $M^{2}_{\mbox{calc}}$ & $\%$ Error & $(D,L^{\pi}_{n})S$ \\
& & & & & & & & & \\
\hline
& & & & & & & & & \\
$\Delta(1232)P_{33}$ & **** & 1.518 & ${3\over 2}^{+}$ & $0^{+},0$ &
${3\over 2}$ & $A_{1}$ & 1.518 & 0 & $(56,0^{+}_{0}){3\over 2}$ \\
& & & & & & & & & \\
$\Delta(1620)S_{31}$ & **** & 2.624 & ${1\over 2}^{-}$ & $1^{-},1$ &
${1\over 2}$ & $E$ & 2.703 &$-$3.0 & $(70,1^{-}_{1}){1\over 2}$ \\
& & & & & & & & & \\
$\Delta(1700)D_{33}$ & **** & 2.890 & ${3\over 2}^{-}$ & $1^{-},1$ &
${1\over 2}$ & $E$ & 2.703 & 6.5 & $(70,1^{-}_{1}){1\over 2}$ \\
& & & & & & & & & \\
$\Delta(1900)S_{31}$ & *** & 3.610 & ${1\over 2}^{-}$ & $1^{-},1$ &
${1\over 2}$ & $E$ & 3.895 &$-$7.9 & \\
& & & & & & & & & \\
$\Delta(1905)F_{35}$ & **** & 3.629 & ${5\over 2}^{+}$ & $2^{+},0$ &
${3\over 2}$ & $A_{1}$ & 3.630 &$-$0.03 & $(56,2^{+}_{2}){3\over 2}$ \\
& & & & & & & & & \\
$\Delta(1910)P_{31}$ & **** & 3.648 & ${1\over 2}^{+}$ & $2^{+},0$ &
${3\over 2}$ & $A_{1}$ & 3.630 & 0.5 & $(56,2^{+}_{2}){3\over 2}$ \\
& & & & & & & & & \\
$\Delta(1920)P_{33}$ & *** & 3.686 & ${3\over 2}^{+}$ & $2^{+},0$ &
${3\over 2}$ & $A_{1}$ & 3.630 & 1.5 & $(56,2^{+}_{2}){3\over 2}$ \\
& & & & & & & & & \\
$\Delta(1930)D_{35}$ & *** & 3.725 & ${5\over 2}^{-}$ & $2^{-},1$ &
${1\over 2}$ & $E$ & 3.759 &$-$0.9 & \\
& & & & & & & & & \\
$\Delta(1950)F_{37}$ & **** & 3.803 & ${7\over 2}^{+}$ & $2^{+},0$ &
${3\over 2}$ & $A_{1}$ & 3.630 & 4.5 & $(56,2^{+}_{2}){3\over 2}$ \\
& & & & & & & & & \\
$\Delta(2420)H_{3,11}$ & **** & 5.856 & ${11\over 2}^{+}$ & $4^{+},0$ &
${3\over 2}$ & $A_{1}$ & 5.742 & 1.9 & $(56,4^{+}_{4}){3\over 2}$ \\
& & & & & & & & & \\
\hline
\end{tabular}
\end{center}
\noindent
Table~4: {\small Oblate top classification of nonstrange baryons
of the $\Delta$ family with $I=3/2$. For further information
see table~3. \normalsize}\\
\vspace{14.5pt}

In table~3 and~4 we show the fit to the nonstrange baryon masses
of the nucleon and the delta family, respectively. These results
obtained in the large $N$ limit are substantiated by exact numerical
calculations. We find a reasonable overall fit for the *** and ****
resonances with an r.m.s. deviation of $\delta_{\mbox{rms}}=0.14$
GeV$^2$. Some characteristic features of the fit are:\\
(i) All experimentally well established resonances are reproduced
by the calculations. Experimentally known but uncertain resonances
with ** status can be accommodated in the fit as well. For example,
for the $\Delta(1600)P_{33}$ resonance with $M^2_{\mbox{exp}}=2.56$
GeV$^2$ we find $M^2_{\mbox{calc}}=2.71$ GeV$^2$.\\
(ii) In the present calculation most states listed in the tables
are rotational members of the ground band ($n_u=0, n_v+n_w=0$).
We have associated $N(1440)$ and $\Delta(1900)$ with $A_1$
vibrational bands ($n_u=1, n_v+n_w=0$), and $N(1710)$ with the
$E$ vibrational band ($n_u=0, n_v+n_w=1$).\\
(iii) The oblate top assignments of orbital angular momentum, spin
and permutation symmetry are similar to the quark model assignment.\\
(iv) The low-lying nucleon resonances, $P_{11}$, $D_{13}$, $S_{11}$,
the cluster of $I=1/2$ resonances in the mass range 1.6-1.7 GeV
and the cluster of $I=3/2$ resonances near 1.9 GeV are well
reproduced. Above this
mass range there are many more resonances predicted than observed
experimentally. This is due to the fact that we have associated the
low-lying $N(1440)$ and $N(1710)$ resonances with vibrational
bandheads. Consequently the rotational states built on top of them
occur low in the mass spectrum. This problem of missing resonances is
known to exist in quark potential models as well.

One possible explanation is that the `missing' resonances indeed do
exist, but that they cannot be resolved individually since there are
many overlapping resonances in that mass region. Another explanation
could be that these resonances are decoupled from the $\pi N$ channel
\cite{KI}. Since at present most experimental information is from
pion scattering, they could simply not be excited in this type of
experiments. Future experiments with electromagnetic probes may
shed more light on this question.
It has also been suggested that the Roper and the $N(1710)$ are
hybrid states \cite{Li}. If that were the case these resonances
are outside the $U(7)$ model space and therefore the coefficients,
$\xi_1$ and $\xi_2$, in the intrinsic part of the mass operator
cannot be determined from the present data. A large value of
$\xi_1$, $\xi_2$ would shift the vibrational bands up in mass,
without affecting the rotational members of the ground band.

Finally, we note that in the $U(7)$ model we have obtained a fit
to the nonstrange baryon masses which is comparable to that in
quark potential models \cite{IK,CI}, although the underlying quark
dynamics is quite different. This shows that the masses alone are not
sufficient to distinguish between different forms of quark dynamics,
{\it e.g.} single-particle {\it vs}. collective motion.
\mbox{}\\
\mbox{}\\
{\bf SUMMARY}\\
\mbox{}

In this contribution we have proposed to use $U(7)$ as a spectrum
generating algebra for a geometry-oriented description of baryons.
It is combined with the spin-flavor and color parts into
a $U(7) \otimes SU_{sf}(6) \otimes SU_{c}(3)$ SGA for baryon
spectroscopy. Although we have limited the discussion to
nonstrange baryons, the $U(7)$ model can accommodate strange
baryons as well. The present model allows one to study
both collective-like and single-particle-like motion in a single
algebraic framework. Collectivity corresponds to coherent and
strongly correlated motion of quarks which requires a strong coupling
of harmonic oscillator shells.
We have shown explicitly that there exists
$U(7)$ mass operators that strongly mix states with different
oscillator quanta, but still preserve the permutation symmetry.

We have applied the model to the family of nucleon and delta
resonances and found good overall agreement with the observed
masses. The fit is of comparable quality to that obtained in
quark potential models. In addition to mass spectra, the model
provides wave functions which can be used to calculate other
observables such as helicity amplitudes and (transition) form factors.
These quantities provide a far more sensitive test to details
in the wave functions than the mass spectrum.
We are in the process of examining a variety of such observables
in the $U(7)$ model. Our goal is to identify signatures which may
distinguish single-particle from collective aspects of quark dynamics
in baryons, as well as to provide guidance to future experiments.
\mbox{}\\
\mbox{}\\
{\bf ACKNOWLEDGEMENTS}\\
\mbox{}

On the occasion of his 50th birthday, it is gratifying and most
proper to dedicate this contribution to F.~Iachello who is actively
involved in advancing the methods and ideas reported in this work.

\newpage
\font\a=cmbx12
\def\refname

\end{document}